\documentstyle[11pt,newpasp,twoside]{article}
\def\edcomment#1{\iffalse\marginpar{\raggedright\sl#1\/}\else\relax\fi}

\def\BP{Ballesteros-Paredes}
\def\'#1{\ifx#1i{\accent"13\i}\else{\accent"13#1}\fi}
\def\VS{V\'azquez-Semadeni}

\reversemarginpar

\begin{document}
\title{Numerical Models of the ISM}
\author{Enrique V\'azquez-Semadeni}
\affil{Instituto de Astronom\'\i a, UNAM, Campus Morelia, Apdo. Postal
3-72, Morelia, Michoac\'an, 58089, MEXICO}

\begin{abstract}
I review recent results from numerical simulations on the
structure and dynamics of the ISM, and attempt to put together a
coherent dynamical scenario. In particular, I discuss results on 1) the
spatial distribution of the gas components, showing that reasonable
agreement between simulations and observations exists, but noting that
in most models the components are simply defined as temperature
intervals, because distinct thermodynamic ``phases'' do not arise; 2)
some statistical issues of the physical fields, like the dependence of 
the one-point statistics of the density field on the effective equation
of state of the gas, the poor correlation of magnetic strength with
density, the energy spectrum in weakly and highly compressible
cases, and the one point statistics of the velocity field; 3) the
effects of spectroscopic observation on distorting the 
physical structures and results from synthetic observations of the
simulations, and 4) several
dynamical and thermodynamical issues, such as the (apparently minor)
role of the thermal 
instability in forming and confining clouds, the continuous, rather
than abrupt, transition between ``phases'', which in turn may be
consequences of the dynamics rather than the agents controlling it,
the possibility of short 
time scales ($\sim $ a few Myr) for molecular cloud formation, and 
the star-gas connection, mentioning that the models generally exhibit
self-propagating star and cluster formation, while the stars
may drive the medium- and small-scale gas motions, and that a ``star
formation instability'' may induce chaotic behavior of the star formation 
rate locally. I conclude with a
round-up view, and a discussion of the work needed ahead.
\end{abstract}

\section{Introduction}

Two of the most influential models of the interstellar medium (ISM)
to date are the two-
and three-phase models of Field, Goldsmith \& Habing (1969) and of
McKee \& Ostriker (1977, hereafter, MO). These models relied on the then
known atomic 
and radiative heating and cooling processes to provide a self-consistent 
picture of the ISM in which the concepts of thermal and pressure
equilibria played a fundamental role. Another important model was the
time-dependent model of the ISM by 
Gerola, Kafatos \& McCray (1974), which was presented as an
alternative to the pressure equilibrium two-phase model of Field et al.\
(1969) but assumed radically different conditions: a constant-density
medium under the influence of stochastic, local heating events that 
should cause strong local fluctuations of pressure and
temperature, because the cooling and recombination times are
comparable or shorter than the time between exposure of a given gas
parcel to one of those heating events. 
Note that the MO model also recognized the existence of
local fluctuations 
in the pressure, although it was still based on the premise of ``rough
pressure balance''.

Nevertheless, both the equilibrium and the time dependent models left
out a number of important aspects in the ISM physics. The multiphase
equilibrium models essentially neglected the possibility of large
pressure fluctuations in the ISM. The time dependent model instead
included this possibility as a fundamental premise, but still neglected
the fact that such fluctuations should
induce motions, which should in general be turbulent and cause strong
density fluctuations. The turbulence involves gas motions at all
scales which not only provide ram ``pressure'', but also mixing, and
can produce compressions rather than ``support'' (e.g., Elmegreen
1993; \BP, \VS\ \& Scalo 1999a). 

Moreover, both the time-dependent and the MO
models omitted other sources of pressure in the ISM, such as magnetic
fields and cosmic rays. The pressure from these agents is in fact 
significantly larger than the thermal pressure (e.g., Boulares \& Cox 1990).
Elmegreen (1991, 1994) has performed a combined instability 
analysis including self-gravity, cooling and heating, and magnetic
fields, but the full nonlinear
behavior can only be dealt with by means of numerical simulations of
the gas dynamics in the Galactic disk. In this review, I will summarize
a variety of results in this area, pioneered by Bania \& Lyon (1980),
concerning the spatial distribution
both on the Galactic plane and perpendicular to it (\S 2),
the one-point and correlation statistics of the
physical fields (\S 3), the comparison between ``synthetic''
observations of the simulations and actual observations of the ISM (\S
4), and several dynamical aspects, including the role of
thermal and ram pressures, the virial balance of clouds, the nonlinear Q 
parameter and the star-gas connection (\S 5). I conclude (\S 6)
with an attempt to present a comprehensive scenario, which should
constitute the first step towards a full dynamical theory of the ISM
that can take over where the time-dependent and the MO models left off.

Given the focus on HI gas of the present conference, this review
concentrates on hydro- and magnetohydrodynamic numerical simulations
of the multi-temperature\footnote{In this review the common term
``multi-phase'' will be avoided and substituted by
``multi-temperature'', since, as discussed in \S 5.1, numerical
simulations do not in general support the existence of sharp phase
transitions in interstellar gas.} gas in the Galactic disk. Thus, the
vast literature existing on numerical simulations of isothermal gas,
aimed at the molecular-cloud regime, will necessarily be
excluded. The reader is directed to the review by \VS\ et al.\ (2000) for 
a discussion of that area current up to 1999, and that by Mac Low \&
Klessen (2002) for the more recent results. Non-hydrodynamical models of 
the dynamical systems kind have also been excluded (see, e.g., the
review by Shore \& Ferrini 1995).

\section{Spatial Distribution of the Gas} 

One of the first concerns of numerical simulations was to show that 
stellar energy injection through ionization heating, is indeed capable
of carving ``holes'' containing hot ($T \sim 10^6$K) or warm ($T
\sim10^4$K) dilute gas, and of producing dense, cool ($T \sim 100$K)
clouds, or in general, what has often been referred to as a
``multiphase'' medium. However, it is important
to note that in all simulations, the temperature spans a continuum of
values, and the so-called ``phases'' are generally defined simply in
terms of ranges within this continuum. Studies
both parallel and perpendicular to the Galactic plane were
conducted. It is also important to note that a general feature of
the dynamic, yet statistically-stationary regime
into which simulations settle, and which is evident from animations,
is that the ``clouds'' have no long-lasting identity, but instead are
continually ``morphing''\footnote{I first encountered this term in the
context of audio production, to denote a continuous
transition from one waveform into another.} (i.e., changing shape,
stretching, merging 
with other structures, and splitting apart; see \S 5.1; Shadmehri, \VS\
\& \BP, this volume).

\subsection{Two-Dimensional Structure on the Galactic Plane}

Early attempts to show the generation of a multi-temperature medium,
even at very low resolutions in two dimensions (2D) parallel to the
Galactic plane, were those by Bania \& Lyon (1980), Chiang 
\& Prendergast (1985), and Chiang \& Bregman (1988). In particular,
Chiang \& Prendergast also identified a ``star-formation'' instability,
similar to a thermal instability (Field 1965; Field et al.\ 1969) around 
an equilibrium state in which star formation balances mass loss, and
stellar heating of the gas balances radiative cooling (see \S 5). 
On the plane,
self-gravity and later magnetic fields, the Coriolis force and shear
were included (\VS, Passot \& Pouquet 1995; Passot, \VS\ \& Pouquet
1995), although still without supernova energy input, and without a
thermal instability-inducing cooling function. Moreover,
\BP\ et al.\ (1999a) suggested that, since clouds are 
really sites of larger-scale converging flows, they should often contain
shocks within them, implying that the magnetic field can develop large
fluctuations and even reversals. Supernova-like energy input at the
highest-density sites was
considered by Gazol-Pati\~no \& Passot (1999), who found a
filling factor of the hot gas at the midplane of typically a few
percent, with occasional excursions of up to $\sim$ 20\%. Gerritsen \&
Icke (1997) gave
the radial profiles and mass fractions of the gas and star surface
density on the plane, finding that the warm gas becomes the dominant gas
component beyond 2 kpc from the center, and that between 2 and 7 kpc the 
mass fractions remain roughly constant, at $\sim 75$\% for the warm gas, 
$\sim$ 15\% for the cold gas, and $\sim 10$\% for what they called the
``lukewarm'' gas, with temperatures intermediate between those of the
cold and warm ``phases''. 

In general, in 2D simulations, the cloud morphology is extremely
filamentary, the filaments arising where large-scale motions collide and 
shock. In turn, ``knots'' form at sites where the filaments collide (see 
also Wada \& Norman 1999, 2001; Chappell \& Scalo 2001). Wada \& Koda
(2001) furthermore find that in simulations with a weak bar-like
gravitational potential, the pc-scale filaments and clumps may organize
into large-scale spirals. Interestingly,
the filamentary morphology appears to persist in the 3D simulations of
Wada (2001). In these simulations, which do not include the magnetic
field, the filamentary structure appears to be the consequence of tidal
interactions between clumps formed by the combined gravitational and
thermal instabilities. Inclusion of the magnetic field may provide
another channel for forming filamentary density structures.

\subsection{Vertical structure}

Perpendicular to the disk, two main paths of numerical study have
been pursued: one assumes that the stellar energy injection maintains a
turbulent, convective regime in which the vertical structure of the
disk is a consequence of both the buoyancy of the hot gas and the bulk
motions induced in the cold medium by the injection of thermal and
kinetic energy by the stellar 
sources. The time average of the instantaneous gas distribution is then
reported as the resulting vertical distribution. Most of the
simulations in this category have neglected 
the magnetic field. The other assumes perturbations about an initial
magneto-hydrostatic equilibrium, as initially proposed by Parker (1966), 
and has studied the nonlinear evolution of the perturbations.

In the first category, Rosen \& Bregman (1995) investigated the
average filling factor and density profile of the various ``phases'' as
a function of height 
above the midplane and of the energy injection rate from
supernovae, finding that only injection rates comparable to the observed 
Galactic rate reproduce the observed vertical distribution of the various 
``phases'', implying that it is the stellar energy injection rate which
is responsible for the vertical distribution. Also, they found that more 
than one exponential component is needed to model the average vertical
density profile, and that their cold gas profile compares favorably to
the observational profile of Dickey and Lockman (1990). Using 3D simulations,
similar results have been obtained more recently by Gerritsen \& Icke
(1997), de Avillez (2000) and Wada (2001). The former authors used SPH
simulations 
of the whole Galactic disk to find the scale heights for the
cold and warm gas and for two stellar populations as a result of the
stellar energy injection, while de Avillez found that, without stellar
injection, the disk collapses vertically to a few tens of pc. Wada
found that the supernova explosions form a plume-like halo.

In the second, magnetostatic category, recent simulations of the
Parker instability in a thick disk have been performed by Santill\'an et
al.\ (2000) in 2D, and by Kim, Ryu \& Jones (2001) and Franco et al.\ (2001)
in 3D. Santill\'an et al.\ and Kim et al.\ concluded that the
Parker instability 
alone cannot be the main formation mechanism of giant molecular clouds in 
the general ISM, because of the low final density enhancements
obtained. However, Franco et al.\ find that the instability can trigger the
formation of such clouds inside spiral arms. The effect of the passage
of a spiral arm has been modeled by
Martos \& Cox (1998), who suggested that it
may induce a ``hydraulic jump'' in which the gas is pushed to
higher altitudes as it it is shocked upon entering the arm and acquires
a higher pressure. Martos et al.\ (1999) have 
furthermore suggested that this jump may trigger star formation at high
altitude over the midplane (over 500 pc). However, the simulations of
Martos \& Cox explored only the isothermal and adiabatic cases (with a
model term for the magnetic pressure -- but see \S 3.1) and did not include
self-gravity, neglecting the possible triggering of thermal 
and gravitational instabilities by the shock compression (c.f.\ \S 5.1).
Exploring this scenario in 
the presence of those agents will be of great interest.

At a more local level, de Avillez \& Mac Low (2001) have shown that
mushroom-shaped structures arise naturally and rather frequently in
their 3D simulations including supernova energy injection, as a
consequence of the buoyancy of the hot gas in supernova remnants, and
proposed that these may correspond to similar structures observed in
recent HI surveys (e.g., Higgs 1999; Taylor 1999)

Three-dimensional simulations addressing the vertical structure of the
Galactic disk including both 
the magnetic field {\it and} the energizing role of the massive stars have
been scarce so far. Korpi et al.\ (1999) placed supernovae randomly in
regions with densities higher than the average. They discussed the
filling factor of the hot gas as a function of height above the plane,
finding values $\sim$ 20--30\% at the midplane, but this
factor is highly sensitive to the degree of spatial correlation of the
model supernova explosions, becoming larger as the supernovae are less
correlated. This probably explains the larger values of the midplane filling
factor they found compared to the values reported by Gazol-Pati\~no \&
Passot (1999), since the latter authors
placed the SNe exclusively at density maxima, making them more strongly
correlated. In the simulations of Korpi et al., the magnetic field is 
initially taken very small, and grows in time, but at the time at which
these authors analyzed their simulations, the magnetic pressure was still
much smaller than the equipartition value with the turbulent component
($\sim 10^{-3}$). Thus, they did not discuss the relative roles
of the stellar energy injection and the magnetic field in maintaining
the disk thickness. This remains an open issue.

\section{Statistics of the Physical Fields}

Given the turbulent, yet statistically stationary nature of the ISM
suggested by the numerical models, it is important to understand the
statistical descriptors. In this section I discuss work related to
various such descriptors.

\subsection{The Mass Density Probability Distribution, the
Density-Magnetic Field Correlation, and the Role of Magnetic Pressure}

The density PDF was found by \VS\ (1994) to have a lognormal shape in
numerical simulations of isothermal flows, which are reasonable
approximations to the flow regime 
in molecular clouds. Passot \& \VS\ (1998; see also \VS\ \& Passot 1999)
proposed that the origin of 
the lognormality is the constancy of the sound speed in isothermal
flows. Indeed, in this case, the probability of a given density jump is
determined simply by the distribution of velocity jumps in the flow,
because the local Mach number, which determines the density jump, is
independent of the local density, and depends only on the velocity
difference across the shock. Thus, all density jumps belong to the same
distribution, and, assuming each jump is independent of the previous and 
later ones (i.e., that shocks are independent from each other), the
Central Limit Theorem can be applied to the logarithm of  
the density, since the density jumps constitute a multiplicative
process. The width of the PDF is proportional to the rms Mach
number. Those authors also found that in the more general case of
polytropic flows, in which $P \propto \rho^{\gamma_{\rm e}}$, where $P$
is the (total) pressure, $\rho$ is the density, and
$\gamma_{\rm e}$ is in general a parameter reflecting the net behavior 
of the pressure with density, the density PDF
approaches a power law at high (resp.\ low) densities for $\gamma_{\rm
e} < 1$ (resp.\ $>1$). They understood this in terms of the dependence
of the sound speed on density, which, when introduced in the
previously-lognormal PDF, gives an asymptotic power law behavior on one
side of the PDF (see Nordlund \& Padoan 1999 for a related discussion).

In more ISM-like simulations in 2D, including magnetic fields, heating
and cooling, the Coriolis force and shear, and localized stellar energy
injection, Scalo et al.\ (1998) and Kritsuk \& Norman (2001) have
reported PDFs with power-law tails at 
high densities, suggestive of a total pressure which behaves as if
having an effective exponent satisfying $0 < \gamma_{\rm e} <
1$, possibly due to the contribution of the magnetic field to the total
pressure. However, in the purely magnetic, isothermal case, several
workers (Padoan \& 
Nordlund 1999; Ostriker, Stone \& Gammie 2001; Passot \& \VS\ 2002) have 
found that for cases with intermediate-to-large Alfv\'enic Mach numbers,
the magnetic field appears quite uncorrelated with the density. Passot
\& \VS\ (2002) suggest that the lack of correlation implies that
the magnetic field does not act as a restoring force, and thus
does not efficiently act as a pressure, instead acting more as a
forcing. This implies that modeling the magnetic pressure by means of a
$\gamma_{\rm e}$ 
may be inadequate in the turbulent case. Padoan \& Nordlund (1999) have
suggested that the lack of 
correlation is a signature of the intermittency of the field, and
pointed out that this is consistent with observations of the magnetic
field strength in molecular clouds. Apparently, the lack of correlation
is also seen observationally (e.g., Crutcher, Heiles \& Troland
2002; Troland \& Heiles, this volume). 

In non-magnetic, multi-temperature simulations, Wada \& Norman (2001)
have reported PDFs with a lognormal tail at high densities, and a
Gaussian one at moderately low densities. Wada (2001, private communication)
has furthermore found a power-law range at very low densities. This
could seem at odds with the results 
of Scalo et al.\ and Kritsuk \& Norman, but upon closer inspection,
there is little 
discrepancy, because most of the PDFs of Wada \& Norman can be
understood in terms of the various polytropic regimes to which the gas
in their simulations is subject. In any case, a detailed study is in order.

The PDF of total projected ({\it column}) density has been discussed by
Ostriker et al.\ (2001), \VS\ \& Garc\'ia (2001) and Burkert \& Mac Low
(2001). Although these works were aimed at isothermal gas, the results
from \VS\ \& Garc\'ia should be readily extendable to the non-isothermal
polytropic case. They suggested that the column density PDF should not
have a unique functional form, but that instead it should transit from
the shape of the underlying 3D density distribution (a lognormal for
isothermal flows) to a Gaussian, as dictated by the Central Limit
Theorem when the contributing sample is large enough. The former (resp.\ 
latter) case occurs in the limit of few (resp.\ many) independent
``density'' events along the line 
of sight (LOS), i.e., when the extension of the observed object along the 
LOS is smaller (resp.\ much larger) than the correlation length. Under
this interpretation, the result of Ostriker et al. that the 
PDF is close to lognormal, suggests that molecular clouds may have
sizes comparable to the turbulent correlation length, a result used by
Burkert \& Mac Low (2001) to suggest that molecular clouds receive their 
energy injection predominantly at the large scales. Observational
results on the column density PDF for the HI gas are in need to shed
light in this direction for this important component of the ISM.

\subsection{The Kinetic Energy Spectrum and the Compressibility of the Gas}

The kinetic energy spectrum is one of
the most extensively discussed properties of turbulent flows.
Simulations of weakly compressible hydrodynamic cases
(e.g., Porter, Pouquet \& Woodward 1992) and of magnetized
incompressible regimes (Cho, Lazarian \& Vishniac 2001) give a spectral
slope close to the Kolmogorov value of $-5/3$. In the magnetic case,
this is consistent with the Goldreich \& Sridhar (1995) theory. However, 
in highly compressible cases, no unique slope has been found, and in
fact it appears to be dependent on the degree of compressibility of the
flow, given by both the rms Mach number and the effective exponent
$\gamma_{\rm e}$. Indeed, for isothermal cases, slopes between $-1.8$
(Padoan \& Nordlund 2000, 3D) and $-2$ (Gammie \& Ostriker 1996, 1+2/3
D) have been found; 
the latter value has also been reported for multi-temperature cases in
2D by Passot et al.\ (1995), and is the one expected for a velocity field
dominated by shocks (e.g., Saffman 1968 [sec.\ 6]; Kadomtev \& Petviashvili
1973). This value is consistent with the observed scaling linewidth-size
relation for molecular clouds (see, e.g., \VS\
1999; \VS\ et al.\ 2000) found by Larson (1981; see also Blitz 1993 for
a summary of more recent results). Observational determinations of the
energy spectrum in the HI gas are often suggestive of spectra closer to
the Kolmogorov value (e.g., Minter \& Spangler 1996; Stanimirovic \&
Lazarian 2001; Dickey et 
al.\ 2001), but sometimes they are not (e.g., Deshpande, Dwarakanath \&
Goss 2000). It is indeed quite possible that the warm neutral and
ionized gas  
exhibit a behavior closer to incompressible Kolmogorov turbulence while
the molecular gas is much more compressible, because the cooling time in 
the former is comparable or larger than the dynamical time, while it is
much shorter in the molecular regime. This implies that the behavior in
the diffuse gas may be
closer to adiabatic (and therefore only weakly compressible)
(Kritsuk \& Norman 2001; S\'anchez-Salcedo, \VS\ \& Gazol 2002).

\subsection{The Velocity Probability Distribution}

An important statistical descriptor that has been discussed mostly in
the context of molecular clouds, but that is in principle equally
important in other regimes in the ISM, in particular the HI gas, is
the PDF of the velocity (vector) field. In the case of
incompressible turbulence, the velocity PDF is known to be nearly
Gaussian, while that of the velocity {\it gradient} (or velocity
difference across positions in a map) is closer to
exponential (e.g., Frisch 1995). Observationally, only
the line-of-sight component of the velocity is available, and only in
projection, and so both line profiles (Falgarone \&
Phillips 1990; Falgarone et al.\ 1994) and PDFs of the line velocity
centroids (Miesch \& Scalo 1995; Lis et al.\ 1996, 1998; 
Miesch, Scalo \& Bally 1999) have been proposed as estimators of the
line-of-sight velocity PDF. Falgarone et al.\ (1994) showed that the
line profiles of weakly compressible turbulence have similar moments to
those of $^{12}$CO and $^{13}$CO observational data, while Lis et al.\
(1998) found that the line centroid {\it difference} PDFs of weakly
compressible turbulence are also similar (non Gaussian) to those of CO
data. From these results, this group has suggested that the regime in
molecular clouds is only weakly compressible, and dominated by vortical
motions.

On the other hand, Miesch \&
Scalo (1995) and Miesch et al.\ (1999) have found that the PDFs of the
line centroid themselves (not of the differences) in both archival HI
data and in molecular-line observations of several star forming regions
are also non-Gaussian, 
with nearly exponential tails, in sharp contrast with the Gaussian PDF
for the velocity characteristic of incompressible or weakly compressible 
turbulence. Miesch et al.\ further pointed out that images of the 
largest-magnitude centroid velocity difference are less filamentary
than expected for weakly compressible cases. From these
results, this group concluded that the turbulence  
in molecular clouds is highly compressible. The rationale here is that
the non-Gaussian velocity increment PDFs would be common to both
the compressible and incompressible cases, and therefore not a good
discriminator between the two, while apparently the non-Gaussian 
centroid PDFs are exclusive of the (proposed to be highly compressible)
molecular cloud 
turbulence. The 2D numerical simulations of a pressureless gas with
(small-scale) stellar energy
injection of Chappell \& Scalo (2001) support this view, giving
exponential velocity PDFs. However, the 3D simulations of self-gravitating, 
isothermal turbulence forced at large scales of Klessen (2000) give
nearly Gaussian velocity centroid PDFs. The reason for this discrepancy
is unclear, and may reside in the different scales of energy injection in
the two sets of simulations, or in the greater compressibility of the
pressureless simulations (which may actually be a better model of the
cool atomic gas in the ISM). Clearly, this remains an open issue, both
numerically and observationally, especially concerning the HI gas in the
ISM.

\section{Synthetic Observations of the Models}

An important application of numerical models of the ISM 
consists of ``observing'' them in a manner similar to how actual
observations of the ISM are performed. Much of the work in this
direction has been oriented towards the molecular cloud regime, and thus 
is out of the scope of this review, but the interested reader is referred to 
the recent work of Padoan, Mac Low, E. Ostriker, Stutzki and
collaborators. Here 
we discuss some applications to the global ISM, and molecular-cloud work
only when it has a direct impact on the global ISM.

One early result found by Burton (1971) and Adler \& Roberts (1992),
albeit somewhat ignored until recently, is that 
structures in position-velocity (PV) space often do not correspond to
actual, connected structures in physical, 3D space, but simply to chance 
superpositions of disconnected structures along the LOS (see also \BP\
et al.\ 1999a; Pichardo et al.\ 2000; Ostriker et al.\ 2001; \BP\ \& Mac
Low 2001; Wada \& Koda 2001). This effect is expected to be especially
important in the HI gas, given that its emmision is mostly optically
thin and widespread. 

Pichardo et al.\ (2000) have recently found two other related
results concerning the structure seen in velocity channel
column density maps. First, as foreseen by Burton (1971),
those authors found that the morphology in the
channel maps shows a somewhat greater resemblance to that of the {\it
LOS-velocity} field than to that of the density field. This result,
together with that of Lazarian \& Pogosyan (2000; see also Lazarian,
Pogosyan \& Esquivel, this volume), that the spectral index of the
``emissivity''  (strictly speaking, of the column density) in velocity
channels of spectroscopic observations depends on {\it both} the spectral
indices of the density and velocity fields,
constitutes strong evidence that the structure of the LOS-velocity field
plays an important role in determining the structure in velocity channel 
maps. The numerical simulations were also used by Lazarian et al.\
(2001) to test the predictions of Lazarian \& Pogosyan (2000),
confirming them to within 10\%.

Second, Pichardo et al.\ found that the ``emissivity'' power spectrum in
velocity channels continues to have substantial power even at scales small
enough that both the density and the velocity spectra have started
to decay, suggesting that the velocity segregation imposed on the
column density by the spectroscopic observation procedure introduces
some spurious small-scale power of its own,
unrelated to the structure existing in the 3D physical space. Henney \&
\VS\ (2002) interpret this as the result of the formation of {\it
caustics} (surfaces of large intensity in PV data cubes) due to the
contribution of finite-extent regions along the LOS to vanishingly thin
velocity channels. This effect may constitute an alternative or
complementary explanation
to the one proposed by Deshpande (2000) for the reported
observations of structures at very small scales (10-100 AU) in HI absorption
studies (see references in Deshpande's paper), which was based on the
suggestion that neighboring lines of sight contain contributions from
even the largest scales (in the LOS direction), which account for the
variability from one LOS to a neighboring one.

A different kind of synthetic observation of numerical simulations was
given by Rosen, Bregman \& Kelson (1996), who compared strip scans of
the integrated ``emission'' (total column density) of cold and hot 
gas in different locations in their simulations to strip scans of HI and
X-ray emission. They concluded that the best matches were provided from
locations in hot bubbles, reinforcing the idea of the Sun being in such
a location. They also found examples of both correlation and
anticorrelation between the HI and X-ray emission, with a slight
statistical preference for anticorrelations.

\section{Dynamical Issues}

One of the main advantages of the numerical simulations is that all the
physical variables are known everywhere in space and at all times. This
allows for a detailed study of the forces acting to create the
structures, the degree to which the structures are transient or bound,
and the feedback effects between the stars and the gas, among many other 
issues. Below, we discuss some of them, after noting that, in general,
the regime suggested by the simulations is analogous to 
Kolmogorov turbulence, in the sense that it is the {\it statistics} of
the flow that remain constant, even though the state is highly
dynamical, and individual structures are transient. Note, however, that
that is probably where the analogy 
ends, as interstellar turbulence is highly compressible (at least in
cool and lukewarm regions), magnetized, and is 
forced over a wide range of scales, rather than only at the largest
scales (see, e.g., Scalo 1987; Norman \& Ferrara 1996; \VS\ et
al. 2000). Moreover, in the compressible case, the turbulent cascade, if
present, is likely to have ``leaks'' from all scales down to the
dissipative scales via the shocks (Kadomtsev \& Petviashvili
1973), contrary to the energy-conserving cascade of incompressible
turbulence.

\subsection{Dynamics and Thermodynamics of the ISM}

One of the fundamental premises of the two- and three-phase models of
Field et al.\ (1969) and MO was that of
pressure balance in the ISM, so that cool, dense HI clouds are confined
by the pressure of their warmer, more diffuse surroundings, even though
in the MO model clouds form in the compresed layers of
expanding supernova remnants. However,
ever since the work of Bania \& Lyon (1980), hydro- and
magnetohydrodynamical numerical simulations have shown that many
features of the ISM, including the thermal distribution, can be
reproduced solely by the action of stellar kinetic energy injection on the
ambient gas. In other words, it is important to determine whether (or
when) the relatively low variability (a few orders of magnitude, compared
to the variations by many dex of the density and temperature) of thermal
pressure is the cause or the effect of the interstellar density and
velocity structure.

A first piece of evidence was provided by the simulations of Chiang 
\& Bregman (1988), Rosen \& Bregman (1995), \VS\ et al.\ (1995) and
Passot et al.\ (1995), 
which contained no thermally-unstable temperature range (although the
Passot et al.\ simulations did contain a constant-pressure temperature
range with thermal $\gamma_{\rm e} = 0$), yet they
produced realistic clouds and intercloud structures that formed as a
consequence of turbulent compressions induced by the stellar energy
input. The main differences with the MO model are that the clouds in the
simulations do not only arise in shells around stellar sources, but
instead the whole medium is in a turbulent state in which compressions
are not necessarily the direct result of an expanding shell, and that,
being formed by dynamical compressions, the clouds are continually
changing shape, merging, getting stretched and disrupted (``morphing''), 
rather than being ``confined'' in any way. This is reflected in
virial-balance analyses of numerical simulations, which show that the
second time derivative of the moment of inertia of clouds in the
simulations, far from being near zero, is generally much larger than the 
contributions from the thermal, kinetic, magnetic and gravitational
energies, and instead has its major contribution from moment of inertia
flux through cloud boundaries (\BP\ \& \VS\ 1997; Shadmehri et al., this 
volume). Of course, simulations including self-gravity (Passot et al.\
1995; Wada \& Norman 1999, 2001) show the existence of 
gravitationally contracting large-scale structures (especially since a
small $\gamma_{\rm e}$ reduces the Jeans length; Tohline, Bodenheimer \& 
Christodolou 1988; Elmegreen 1991; \VS,
Passot \& Pouquet 1996), as well as of collapsing
small-scale ones. Meanwhile, the internal structure of the giant cloud
complexes is continually reshaped by the stellar energy injection. One
could argue that the clouds are confined by turbulent 
pressure, but this would be misleading, as turbulence contains
large-scale chaotic motions (at the scale of the whole cloud) which involve the
distortion, and in general, transient character of the clouds (\BP\ et al.\
1999a).

An important consequence of this turbulent scenario is that it raises
the possibility of molecular clouds having formation times (of order a
few Myr) shorter by factors of at least half an order of magnitude than
previous estimates (e.g., Blitz \& Shu 1980), assuming they form from
turbulent accumulation of neutral gas (\BP, Hartmann \& \VS\ 1999b;
Elmegreen 2000). In the scenario of \BP\ et al.\ (1999b), molecular
clouds form from larger-scale converging flows 
in the HI gas that therefore have larger velocity dispersions than those 
normally associated to the molecular gas (because larger scales have
larger velocity dispersions in most turbulent flows). A key issue,
however, is that most of the accumulation process may occur in the
atomic phase, and only when the column density has reached high enough
values does the molecular phase appear (Hartmann, \BP\ \& Bergin
2001), eliminating concerns that in such short times not much
accumulation can be achieved unless the density of the initial medium 
is already (too) large (Pringle, Allen \& Lubow 2001). The short
``formation'' time scales refer only to the time after molecular gas
appears. 

In this dynamic scenario, the relatively weak variability of the thermal
pressure in the global ISM is simply a consequence of cooling and
heating functions that 
imply a slow variation of the thermal-equilibrium pressure with
density, while the density field is driven by the turbulence, which in
turn is powered by the presence of many local heating stellar
sources. Indeed, \VS\ et al.\ (1996) noticed that, if the cooling times
are short compared to the dynamical times and the cooling can be
approximated as a (possibly piece-wise) power law, the gas follows
a nearly polytropic behavior in which $P
\propto \rho^{\gamma_{\rm e}}$ for a power-law cooling function, with
$0 \la \gamma_{\rm e} \la 1$ for $100 \la T/{\rm K} \la 8000$ (Scalo et
al.\ 1998). Note
that a thermally unstable range would have $\gamma_{\rm e} < 0$. Certainly,
the functional form of the pressure feeds back on the flow,
but mainly to determine the response of the density field to the
turbulent compressions, as evidenced by the dependence of density the
PDF on $\gamma_{\rm e}$ (Passot \& \VS\ 1998; see \S 3.1). Of course,
the pressure near stellar injection sites is much larger than that on the 
average ISM, as essentially they may be regarded as either lying on a different
equilibrium between the cooling and the local stellar heating (which
is much larger than the diffuse background heating), or completely outside of
thermal 
equilibrium, if the dynamical times are short enough. In the simulations 
of Passot et al.\ (1995), which included only OB-star ionization-like
heating, the pressure in ``HII'' regions was $\sim 10$ times larger than 
the average. The simulations of Mac Low et al.\ (2001), which included
supernova energy input, exhibited pressure variations of up to three
orders of magnitude. However, the stellar heating is a relatively local
phenomenon, which 
serves mostly as the forcing for the global ISM (Avila-Reese \& \VS\
2001), while the large fraction of the volume that is not under the
direct influence of the sources has a pressure determined
essentially by the turbulent density fluctuations through the applicable
cooling and heating laws.

On the other hand, ISM simulations including a thermally unstable
temperature range between the cold and warm ``phases'' show density PDFs
without clear signs of bimodality,
as would be expected for actual phase segregation (\VS, Gazol \& Scalo
2000; Wada \& Norman 2001; Kritsuk \& Norman 2001), and temperature PDFs
which, although bimodal, contain significant
amounts of gas in the thermally unstable range (Gazol et al.\ 2001;
Kritsuk \& Norman 2001; see also Gerritsen \& Icke 1997);
the same effect is seen in the plots of Korpi et al.\ (1999), although
they did not discuss it. The multimodality of the temperature PDF does not 
necessarily signal the existence of phases, but simply of heating and cooling
processes which favor certain values of the temperature as a function of 
density. The simulations of Korpi et al.\ (1999), which include
supernova heating, and thus hot ($\sim 10^6$ K) gas, have
roughly flat density PDFs except for a pronounced peak at the density of 
the hot gas, indicative of the relatively large relative filling factor
of this temperature range, but still do not show significant signs of
``phase'' segregation. 
For there to exist real phase segregation, a discontinuity must
occur in one or more of the thermodynamic variables. However, in the
simulations there are no such discontinuities (within the numerical
resolution and smoothing of discontinuities), and instead the flow
appears as a continuum, with the only discontinuities being dynamical
shocks. An exception is the case of the simulations by Hennebelle \&
P\'erault (1999, 2000) and Koyama \& Inutsuka (2000), who studied the
triggering of thermal instability by shocks passing through initially
stable gas. However, these authors did not consider the case of a
fully turbulent ISM in the presence of other forces besides the thermal
pressure gradient. Instead, although further testing at higher
resolution is still necessary, the simulations of the fully turbulent
ISM strongly suggest that the ``unstable'' range is 
actually quite populated in the ISM. A number of observational works
point in the same direction (e.g., Dickey, Salpeter \& Terzian 1977;
Heiles 2001, this volume), although these require further confirmation 
themselves (see Miville-Deschenes, Hennebelle \& P\'erault, this volume, 
for an opposing view). 

The mechanism responsible for the at least partial inhibition of the
thermal instability is not quite understood so far. Magnetic pressure
possibly plays a role (but see \S 3.1), and velocity fluctuations may
inhibit the instability if they have shorter time scales than 
the cooling time, a phenomenon which is most likely to occur at small
spatial scales and/or low densities (S\'anchez-Salcedo, \VS\ \& Gazol 2002).
It is interesting to note that a similar result has been recently
reported for the Parker instability, which apparently can be partially
or completely suppressed by the presence of fluctuations in the magnetic
field (Kim \& Ryu 2001).
Also, simple continuous recycling from one phase to another may imply
the existence of a persistent population of gas traversing the unstable
regime, as suggested by Lioure \& Chi\`eze (1990).

A complementary point of view is that of Wada,
Spaans \& Kim (2000), who have recently proposed that the pure
gravitational instability, aided by thermal instability, is capable of
the formation of large low density regions (``voids'') in the global
ISM, which, in their picture, correspond to (some of) the observed HI
holes in galactic disks. This is 
not inconsistent with the previous view of shaping the interiors of
giant cloud complexes through the turbulence powered by stellar energy
injection. Indeed, the largest cloud complexes may be gathered together
by large-scale processes such as gravitational instabilities or the
passage of spiral density waves, necessarily leaving behind, by simple
mass conservation, large cavities which in fact must occupy large
volumes, due to their lower densities. On the other hand, the stellar
energy input, which is most likely dissipated within 
relatively short distances from the injection sites (Avila-Reese \& \VS\
2001), can have mainly the role of shaping the complexes'
interiors. Wada et al.\ distiguished between stellar-carved cavities,
which should be filled with hot gas, and instability-carved ones, which
should be filled mostly with warm gas.

The nonlinear development of gravitational instability in magnetized
galactic disks has been recently studied by Kim \& Ostriker
(2001a,b). These authors note that the amplification of non-axisymmetric
perturbations in the presence of shear saturates in linear theory, and
that thus any $Q$ threshold for nonaxisymmetric gravitational runaway
must originate from nonlinear effects. In this context, they numerically 
estimate the threshold values of $Q$ in the nonlinear case. They also
distinguish between the  
mechanisms operating behind the modified swing amplification and the
magneto-Jeans instability.

An interesting point to note is that even in works without stellar
energy injection, the outcome of the development of the various
instabilities is often a chaotic, or turbulent, medium (Wada et al.\
2000; Kim, Ryu \& Jones 2001; Kim \& Ostriker 2001a,b; Koyama \&
Inutsuka 2001; Kritsuk \& Norman 2001), and so these large-scale
instabilities can also be sources of the internal turbulence of the cloud
complexes (see also Sellwood \& Balbus 1999 for an
analytic discussion of the magnetorotational instability powering
turbulence in HI disks).

\subsection{The Star-Gas Connection}

In general, most simulations including non-random prescriptions for star
formation 
(SF) show that it self-propagates and, when other agents such as
self-gravity and cooling are also included, spontaneous SF
occurs too (e.g., Chiang \& Prendergast 1985; \VS\ et
al.\ 1995). Moreover, as has been noted repeatedly, the stars in the
simulations provide the energy feeding the gas motions, and a feedback
cycle, analogous to that of Oort (1954), is established. However, the
cycle is highly chaotic locally (\VS\ et al. 1995, Passot et al.\
1995; Gazol-Pati\~no \& Passot 1999; Wada \& Norman 2001; see also Shore 
\& Ferrini 1995), and depends on
global quantities such as the mean magnetic field strength, and on
stellar properties such as the energy deposited per stellar source
(Passot et al.\ 1995; Gerritsen \& Icke 1997; Gazol-Pati\~no \& Passot
1999). Other properties 
of the SF process, such as the slope of the two-point correlation
function describing the clustering of young stars, can also be accounted for
by this type of numerical models (Scalo
\& Chappell 1999). Additionally, Chappell \& Scalo (2001) have studied
the ratio of present to past-average SF rate as a function of various
parameters in their simulations, finding, among other results, that
starbursts can only occur when the past-average rate is low or the
system is small, that the broad distribution of this ratio in
late-type systems can be understood as a result of either a small size
or a small metallicity, which imply that larger expanding shell column
densities are required for gravitational instability, and that
exponential tails in the velocity distribution are due to multiple shell 
interactions, not individual stellar winds.

It is worth noting here that the ``star formation instability''
identified by Chiang \& Prendergast (1985), which was one of the few
attempts to investigate analytically the star-gas interplay within a
hydrodynamic approach\footnote{Many attempts have been made in the
context of the dynamical systems approach to ISM models; see, e.g.,
Shore \& Ferrini 1995}, has not been discussed in recent years, even 
though its effect may be present in all models with self-consistent
prescriptions for star formation, possibly explaining its chaotic
evolution. Another important connection that has not
been considered in numerical models is the excitation of density and
velocity perturbations in the gas by the gravitational effect of the
stars (Kegel \& V\"olk 1983). These issues certainly deserve further
investigation.

\section{Conclusions: the Emerging Scenario and the Work Ahead}

In this review I have summarized a large body of results derived mainly 
from numerical simulations of the ISM, which have allowed workers to
capture many of the complex dynamical aspects of ISM structure and
evolution. The numerical models have suggested a 
much more complex ISM than a simple three-phase medium in pressure
equilibrium. Instead, a turbulent continuum has emerged, apparently
without sharp phase segregation, and in which turbulent ram pressure is
mostly responsible for cloud formation within the large complexes, while
strong thermal pressure imbalances due to local stellar energy injection are
responsible for powering the turbulence through expanding
shells, although otherwise the near constancy of the thermal pressure has
little effect in confining density structures, since {\it the absence of a
pressure gradient does not imply that inertial motions cannot exist}. The
medium is not simply a collection of overlapping 
shells with clouds forming in their compressed layers, but instead is
globally turbulent, with shells and all structures in general
``morphing'' and  merging into the global turbulence. At the largest scales,
combined gravitational, magnetic and thermal instabilities appear
to contribute, together with spiral density waves and supershells, to the
formation of the largest complexes and perhaps voids, as well as
possibly feeding the turbulence from the largest scales.

The individual structures being transient, the appropriate description
for the global structure is statistical. Some statistical measures of the flow
(the density PDF and the energy spectrum) depend on the cooling ability
of the flow, which in turn determines its compressibility. Since this in 
turn determines the ability to form gravitationally bound structures
through compressions, and therefore stars which feed the turbulence, a
full feedback loop clearly exists analogous to the old Oort (1954) cycle.

However, as this review probably makes evident, the body of results is
highly scattered, and a coherent, global theory of the ISM is still
lacking. Such a theory should be able to predict fundamental statistical 
indicators of the ISM such as its topological properties (i.e., the
statistical distribution of the mass and other physical
quantities), the rate of production of collapsing objects (the star
formation rate and its efficiency) and the mixing rate of the processed
chemical elements, as 
well as the thermal and radiative properties of the various temperature
and density regimes, all as a function of simple input physics, such  as 
the total available mass and angular momentum, the atomic processes
determining the cooling rates, and the energy 
injection per source. Note that, in
principle, even ``parameters'' such as the energy injection rate and the 
scale of injection, should be derivable from the theory (or model),
because the sources in this case are linked to the 
flow dynamics, as is the case of large-scale instabilities and of the
star formation rate. Many
of the physical variables are expected to be highly fluctuating  
locally, and the prediction of both average (see, e.g., Blitz, this
volume) and typical fluctuation values is crucial. Simulations
including all relevant physical ingredients (self-gravity, disk
rotation, magnetic fields, cosmic rays, stellar energy input, spiral
density waves, chemistry) and that can perform the necessary radiative
transfer, all at high enough resolution, are needed. So are careful
experiments which allow disentangling the effects of all those
ingredients, and of the average quantities versus those of the
fluctuations. Consideration of the gravitational effect of the stars is
also needed, which implies that hybrid gas+stars similations are
needed. Then, continuous feedback to/from observations is essential through
the production of synthetic observations of the numerical models. A
long, exciting way still lies ahead to a comprehensive theory that
grasps the dynamics, thermodynamics and statistics of the ISM with the
same level of detail with which the time-dependent and the multi-phase
models grasped the thermal and radiative issues.

\acknowledgements

I gratefully acknowledge Pepe Franco, Alex Lazarian, Marco Martos,
Thierry Passot and 
Keiichi Wada for a critical reading of the manuscript and useful 
comments and precisions. Very specially, I thank John Scalo, who
pointed out
various important topics, and helped in making sense of some others. This
work has received partial financial support from CONACYT grant 27752-E
to the author, and has made extensive use of NASA's ADS service.

\end{document}